# Investigating the potential impact of values on requirements and software engineering

Alistair Sutcliffe, Pete Sawyer, Wei Liu and Nelly Bencomo
School of Engineering and Applied Science, Aston University, Birmingham B4 7ET, UK

*Abstract*--**This paper describes an investigation into value-based software engineering and proposes a comprehensive value taxonomy with interpretation of design feature implications. The value taxonomy is used to assess the design of Covid-19 symptom tracker applications, contrasting the UK's NHS phase 1 and 2 designs which adopted centralized, then decentralized, architectures. The value/ feature analysis is also applied to the King's/Zoe Covid app which does not detect proximity, instead relying on user self-reporting. Value analysis illuminated design choices but was insufficient to account for download acceptance of the apps. We argue that motivational cost-benefit analysis needs to complement a values-based approach.**

*Index terms*--**values, motivation, design features, technology acceptance**

## Introduction

Values and ethical concerns have been investigated in software engineering (SE), with particular emphasis on sustainability and privacy [1-2]. While value manifestos [3-4] have drawn attention to values and related phenomena, guidance to inform software engineers about potential design implications of values has been less forthcoming. Values may emerge from ethnographic analysis [5-6], although interpreting their design implications relies on the intuition and experience of the analyst. Value-Sensitive Design (VSD) [7] proposes a scenario-based probing of values in the requirements phase to sensitize designs to stakeholder values; while Value-Based Requirements Engineering (VBRE) [8-9] also follows a scenario-based analysis with a taxonomy of values and motivations as an elicitation guide, supplemented with some hints about potential design implications.

In spite of these initiatives, the state of the art has not progressed beyond case study exemplars investigating value implications in a single domain/application context [2, 10]. Our goal is to develop more informative guidance for applying value-based analysis in SE, by connecting values to their implications in design features. Technology Acceptance Modeling [11-12] has demonstrated the importance of values in determining actual use of IT products, with other variables such as user motivation. This raises questions concerning the sufficiency of values for explaining system acceptance. These aims are summarized in two research questions:

1. What are the software design implications of values?
2. To what extent do values contribute to system acceptance?

In following sections of this paper, we first review research on values and related concepts; this is followed by proposing a guiding framework for considering the implications of values in design and system acceptance. We then apply our framework and process to a case study of value-based feature assessment and design trade-offs focused on current track-and-trace applications that inform users of exposure risks to Covid-19 infection.

## Related Research

A taxonomy of social and political issues with guidelines for recognizing affective reactions among stakeholders [13] was applied to analyzing requirements for ERP applications. In an application of Activity Theory to SE, [14] elaborated UML schema for social issues and proposed patterns for recognizing stakeholder conflicts; however, they did not give specific advice about eliciting or analyzing users' values and emotions. User values have been analyzed as cultural attributes such as power distance and individualism [15]; however, apart from isolated examples (e.g. [16]), few reports of the application of value-based cultural implications have emerged.

In human-computer interaction (HCI), the value-sensitive design method [7] provides a process for eliciting user values with scenarios and storyboarding techniques. However, VSD does not focus directly on specification; instead, it aims to elicit users' attitudes and feelings about products and prototypes. Values and affective responses have been investigated by [17] in worth maps, which attempt to document stakeholders' views about products or prototypes expressed in stakeholders' language as feelings, values and attitudes.

[3] described a research road map for building an ontology of values, operationalizing implications of values in software and resolving value clashes between developers and stakeholders. [4] demonstrate the use of values in agile development and participatory design using the Schwartz [18] taxonomy (see below). This taxonomy was applied in a case study deconstructing the European data privacy regulations (GDPR) [2] showing the direct and indirect links between rights, principles, and values such as privacy, trust, transparency, accuracy and lawfulness. The Schwartz values in combination with Holbrook's [19] consumer values were applied to requirements modeling using i* strategic dependency models to elaborate the implications of values as soft goals for requirements in distance education applications [20]. This study

illustrated how values such as 'fun' and 'ethics' could inform requirements for video media interaction and cheating detection; however, the proposed method relied on analyst expertise in interpreting value-related implications. While the Schwartz taxonomy has been the most frequent choice for application in SE, more comprehensive value taxonomies have been reported [21].

In conclusion, the importance of values and social issues has been recognized in SE; however, although the connection between values and high-level system architecture was proposed by [22] in the form of connections between values such as 'freedom' and 'autonomy' and system-initiated control functions, advice on how to elicit and deal with the design implications is scarce, and few validated methods have been reported.

APPLYING VALUE TAXONOMIES IN SE

Values are language-bound psychological constructs which do not lend themselves to precise definition. Values appear to be tacit knowledge: we recognize them when we encounter them but trying to articulate them beforehand is difficult. In psychology values are related to other constructs such as intent (goals), attitudes, and beliefs. As noted by researchers in ethnography, such knowledge is contextualized and can only be understood by reference to a situation [5, 6]. The key to understanding values is through language with reference to existing knowledge held by the individual; for example, sustainability may be understood in terms of related concepts: green issues, environmentalism, climate change, ecological concerns. Probably the most influential value theory was created by Schwartz [18] who defined values as "guiding principles in the life of a person or group". His taxonomy of ten basic values is organized into four higher-order motivational categories: openness to change, self-enhancement, conservation and self-transcendence. Motivation theory, which preceded Schwartz's taxonomy, proposed six need-based categories which intersect with Schwartz's values: physiological needs, safety needs, social belonging, self-esteem, self-actualization, and transcendence. Some values, e.g. broad mindedness, may also be related to personality traits which are measurable characteristics of people such as in the OCEAN taxonomy (openness, conscientiousness, extroversion, agreeableness and neuroticism). We adopt the meta-taxonomy of [23], which is the most recent review of value concepts, and augmenting it with morality and ethics [8, 23] which were not explicit in the taxonomy but may be cross-referenced to honesty and justice. Aesthetics [9] is added as the human attitude to beauty, while emergent values, e.g. sustainability/ greenness, are acknowledged as a placeholder category among social values where new values are more likely to emerge.

To progress connecting values to design, first we investigate the implications of values for software requirements, including the interaction between humans and software, i.e. an emergent property of socio-technical systems. The implications were generated by a consensus discussion among the authors, whose suggestions were then merged into an agreed list. Implications were also drawn from the VBRE method [8, 9]. The values are ordered in four sub-divisions: personal values closely related to motivation; values closely related to personality traits; personal characteristics which are measurable properties of individuals; and societally oriented values. These categories illustrate the connection between values and other psychological phenomena and are not part of the meta-taxonomy [23]. Table 1 summarizes the possible implications for software design and HCI requirements as functional allocation (partitioning responsibilities between manual operation or automation [24]) and emergent socio-technical systems requirements for information, decision support, etc.

*A. Personal-motivational values*

*Wealth* may influence design since software systems have a monetary value for stakeholders-owners. Wealth may impact on project management, in planning development projects where the monetary value of a new system, or potential 'worth' is estimated, with cost-benefit analysis of tangible (monetized) and intangible benefits. *Accomplishment/achievement* is important in most interactive systems where fulfilling goals is a joint human-computer venture. The more the system design actively supports human goals (see also helpfulness), the greater is its contribution to achievement. *Self-respect* reflects self-confidence and positive evaluation of oneself as self-esteem. Some implications are providing means for explicit endorsement by accumulation of followers, likes, etc. in social media or virtual reality systems that include self-representation via avatars. Other implications may be indirect, for instance, when exposing oneself online leads to trolling, or where automation de-skills a human activity leading to loss of employment and consequent poor self-esteem. *Security-safety* values have a clear design implication for privacy/security and applications involving sensitive data, as well as in the many computerized control systems with safety and reliability engineering requirements for hazard detection and error prevention.

*B. Personality-related values*

*Helpfulness* has implications in design of interactive features which support human activity, e.g. help systems, guides, question-answering and explanation facilities which aim to help people operate machines and achieve their goals. *Broad-mindedness* includes tolerance and adaptability with design implications for configurable, customizable designs that respond to the differing needs of individual users, groups, and cultures.

*C. Personal characteristics values*

*Honesty* implies requirements for explaining the provenance and transparency of designs, with explanation of system actions. Honesty has an indirect impact via trust in software systems, which is manifest as non-functional requirements (NFRs) for predictability, transparency and visibility of system actions. *Creativity* is manifest in end-user development, software reuse, component-based SE, open-source software, and product lines, where these environments support human creative design [25]. *Intelligence* clearly applies to people, and to machines in the trivial sense of whether they are considered intelligent (i.e. AI) and logical (all computers). *Responsibility* is related to trust and competence which may be modeled as dependency relationships in i* [26-27] or in the goals-skills-preferences framework [28]. *Aesthetics* influences user interface design directly through user experience [29], but may also be assessed for the physical presence, appearance, and actions of human representations in software systems, e.g. avatars, VR (Virtual Reality), robots.

*D. Social values*

*Social order* is a heterogeneous category which includes organizational stability, respect for the law, and peace. Some implications apply to social media which may have a demonstrable negative effect on social order, while the growth of AI may become a threat to social order. *Freedom* has implications for design through autonomy, surveillance and control. For example, awareness requirements [30-32] concern design decisions about the degree of surveillance, data capture and system functions where the system may control human behavior and decision making. *Equality* belongs with honesty as aspects of morality and ethics. The concept of fairness has an indirect effect on design of AI systems which have the power to classify and arbitrate between individuals or groups of people [33]. *Justice* is related to fairness, which is a growing concern with the impact of AI systems where recommendations and classifications are criticized for being unfair and biased [10, 33], consequently influencing human trust in technology.

Table 1. Value implications for software engineering

| Value | Notes on impact | Design Implications |
|---|---|---|
| *Personal motivation values* | | |
| Wealth | Indirect effect cost-benefit analysis, project management | Gaming incentives |
| Accomplishment/ achievement | Indirect effect on interactive system design via software contribution to joint goals | Functional allocation, socio-technical systems |
| Self-respect/ esteem | Implications for self-representation on social media and VEs. Indirect effect on human work | Functional allocation, socio-technical systems |
| Security/safety | Personal data, privacy, control systems, safety engineering, hazard-threat-mitigation analysis | Safety & privacy requirements |
| *Personality- related values* | | |
| Broad-mindedness | Implications for adaptability, adaptive systems, customizable and configurable designs | Configuration facilities, software versions |
| Helpfulness | Altruistic, motivation to benefit others. Affects design of any interactive system as well as help facilities, explanation, etc. | User support requirements, avatars, agent behaviour |
| *Personal characteristics* | | |
| Creativity | Extends adaptability with facilities for users to compose and create system variants | Component-based SE, software reuse |
| Honesty/integrity | Indirect effect via trust, transparency, and explanation | Explanation facilities, visualization |
| Intelligence/ wisdom | Human characteristic, few design implications | Functional allocation |
| Responsibility | Related to trust, modeled as means-end to dependency relationships | Functional allocation |
| Competence/ capability | Skills, abilities, experience of people, functional capabilities of machines | Functional allocation |
| Aesthetics | Form and function of products, human attributes and judgment | UI design, UX product design, avatars |
| *Social values* | | |
| Social order | Human desire for stability, social media/AI trends | Network effects STS |
| Freedom/ autonomy | Increasing autonomy of AI; autonomy for people is affected by monitoring and surveillance | Awareness requirements, AI/ML, STS |
| Equality/fairness | Implications for AI/ML training and recommender systems | AI/ML training, bias detection |
| Justice/fairness | Implications for AI/ML training and recommender systems | AI/ML training, bias detection, STS |

CASE STUDY: VALUE ANALYSIS OF COVID-19 RISK ADVISOR APPLICATIONS

Covid-19 contact alerting-tracing systems have involved a choice between decentralized architectures developed by Apple and Google, and centralized architectures [34-35]. The UK National Health Service (NHS) first adopted a centralized architecture in May 2020 which was subsequently abandoned. In September 2020, a decentralized architecture based on Google/Apple technology was adopted and this version has been released for general use. Both case studies are retrospective in nature, using values to evaluate design feature implications as well as investigating possible reasons for technology acceptance considering the possible value impacts user downloading and use of the Covid-19 track-and-trace systems.

*A. NHS Mobile phone proximity-monitoring apps*

Both centralized and decentralized architectures use features of mobile phone operating systems to detect proximity of other users [34-36]. If the user has tested positive for Covid-19 and reported their test to the system, the history of recent contacts is interrogated, and warning messages are sent to the owners of those phones urging them to self-isolate. In a centralized system, the personal details of the user and all the recent proximal contacts are sent to a central database. In a decentralized system, personal and contact data are not exchanged; only a warning is sent to recent proximal contacts [37]. Currently, both centralized and decentralized apps offer little advice to the user to help with Covid-19 symptom self-management.

The NHS vn2 App is illustrated in Figures 1(a) for the general interface and 1(b) showing risks of Covid-19 infection in the user's local area.

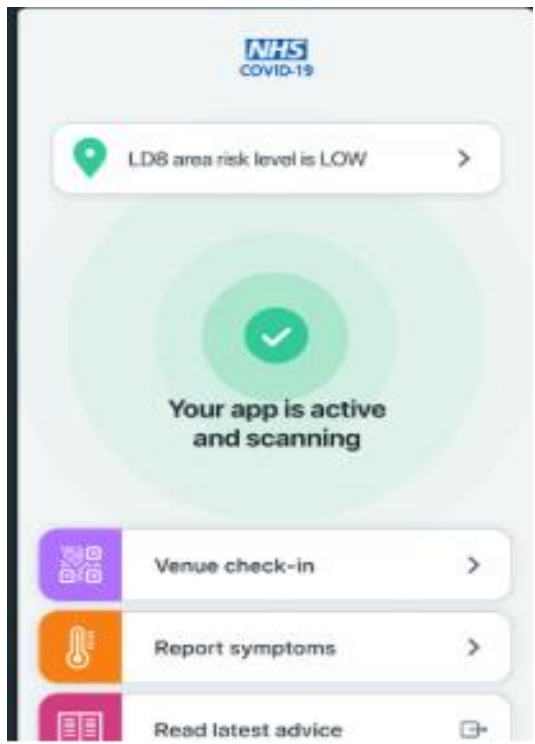

Fig. 1(a). Illustration of the main user interface of NHS vn2 (decentralized)

The user interface of the vn2 App was similar. Other features are limited information about symptoms, symptom reporting, and checking risk by venue, although this is dependent on display of a QR code in the restaurant, pub or shop being visited. The App has limited integration with NHS services; only on-line booking of a test is provided, and there is no integration with supporting information on the NHS website. The mapping of design features to values is illustrated in Table 2.

Negative assessment of values is assigned either because a design feature is absent, e.g. system management advice is absent but could be helpful, or because the design feature may adversely effect users' concerns for that value, e.g. central database may lead to data loss, security and privacy fears. A neutral assignment denotes that the design feature is present, but it makes little or no contribution to the value, e.g. system explanation is present, but it is either limited or so obscure that it can not be considered to be of much use. The negative effectiveness rating is a consequence of poor download rates (to date) of both versions, which limits their effectiveness in Covid-19 infection tracing. As may be expected, security/privacy values have the most salient impact, followed by helpfulness for the individual and society, also expressed as the altruism motivation. Secure data encryption has been emphasized in both versions as a positive contribution to counter privacy fears. The centralized design has more negative security/privacy implications than the decentralized design; however, this may be balanced by benefits of more targeted risk analysis and better integration with scientific advice supported by centralized data. Privacy fears in the centralized version may spill over into a negative assessment of freedom/autonomy from 'big brother' state control.

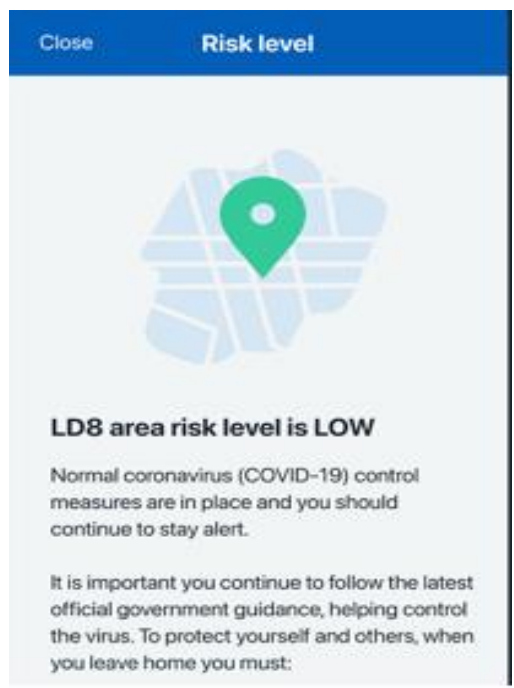

Fig. 1(b). Risk alert map

Table 2. Value implications of the NHS vn1 (centralized) and vn2 (decentralized) design features. *Negative* impacts are coded in *underlined italics*, neutral impacts in regular font and positive in **bold**.

| Design Feature | Centralized | Decentralized |
| --- | --- | --- |
| Download / usage choice | *Freedom/ autonomy* | Freedom/ autonomy |
| Personal identity known | *Security/ privacy* | Security/ privacy |
| Phone / device identity known | *Security/ privacy* | *Security/ privacy* |
| Central registry / database | *Security/ privacy* | Security/ privacy |
| Contactee ID known | *Security/ privacy* | Security/ privacy |
| Contactee risk analysis | **Helpfulness** | **Helpfulness** |
| Secure data encryption | **Security privacy** | **Security/ privacy** |
| System explanation | Honesty | Honesty |
| Treatment follow-on service | Helpfulness | Helpfulness |
| Symptoms management advice | Helpfulness | Helpfulness |
| Integration with scientific research | Altruism | *Altruism* |
| Effectiveness in disease control | *Helpfulness* | *Helpfulness* |

Since the Covid-19 pandemic is a society-level threat, society values should be important. Although social order, justice, and freedom may be threatened by the pandemic, these values have indirect connections to design. More direct design implications are associated with the freedom/autonomy, honesty, helpfulness group of values which are associated with trust, where the NHS brand, which has an excellent reputation in the UK, should motivate downloading and use. Monitoring contacts with others poses a potential threat to privacy and security of personal data, which may be misused by authorities. Honesty is associated with explanation of system aims as well as storing depersonalized data which may mitigate security fears. Both versions included some limited explanation of how the system worked. The centralized version has five negative value scores, invoked by security concerns, possibly compensated by altruism encouraged by better data provision for scientific research. The decentralized vn2 is rated as neutral for the same security/privacy features, as users' perception of a positive contribution to personal data privacy is likely to need further evidence of data security in use. Overall, the feature analysis favors the decentralized design, although the contribution to helpfulness by both versions is limited since explanation was limited and there is little information to help with symptom management beyond the advice to self-isolate and get a test.

Acceptance of monitoring apps may depend on individuals' assessment of the trade-off between their cooperative altruistic motivation and the fear of loss of control of personal data linked to the security fears. Decentralized systems may mitigate security fears; however, individuals with altruistic motivation might be persuaded to use centralized systems if security safeguards are implemented and clearly explained.

Experience with the first UK NHS App (vn1) [38] indicated that the percentage download/uptake was poor: ~ 20% against a target of 60%; furthermore, users' attitudes were not positive. The reasons for poor usage and lack of success in detecting contacts appear to be a mix of technical problems with Bluetooth-related proximity detection and lack of user acceptance [36, 39]. Poor trust in the government-sponsored app, even though it had the NHS brand, and lack of publicity about the benefits (i.e. helpfulness), may have contributed to the failure. The UK then changed to the decentralized model for a second prototype, launched in September 2020. Initial reports noted improved downloads (~ 35%) but many usability problems [40]. The value analysis suggested that the decentralized vn2 app should be more acceptable; however, longer-term success is not immediately apparent from the values analysis. Values may need to be considered in the context of a cost-benefit argument for technology acceptance, e.g. does altruism outweigh security fears for most users?

*B. King's/Zoe symptom-reporting app*

The King's College/Zoe Covid-19 symptom tracker app [41] (see Figures 2(a) and (b)), does not have proximity sensing; instead, it asks the user a few health-related questions. If the answers indicate a Covid-19 infection, the app advises taking a confirmatory test and self-isolation. The app requests daily reporting of user symptoms (if any) and therefore the cost of use is mildly intrusive; however, it does provide a more comprehensive information service as well as local area infection maps for users to assess their own infection risk, features supporting helpfulness. Data is centralized; hence, users who subscribe to the app have to evaluate the trade-off between the fear of data loss and privacy and the benefit of contributing to research and tracking the virus as a warning for others (altruism and helpfulness).

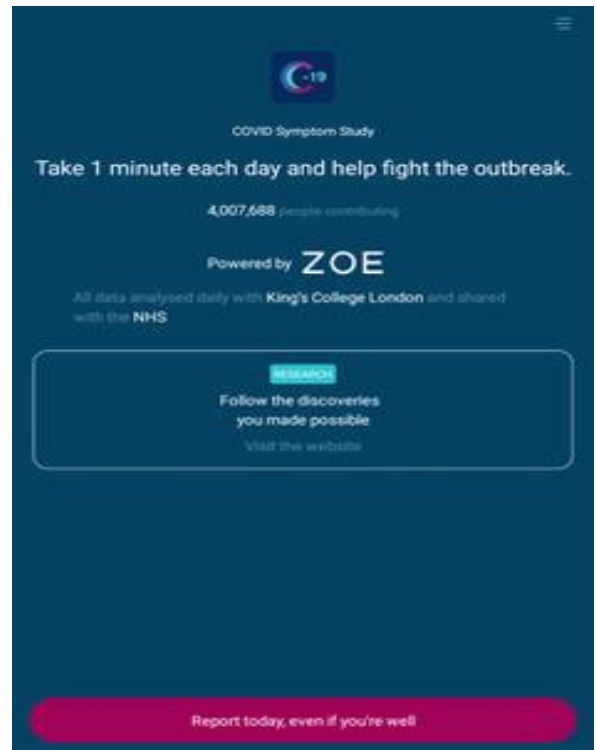

Fig. 2a. Illustration of the main user interface of King's/Zoe app

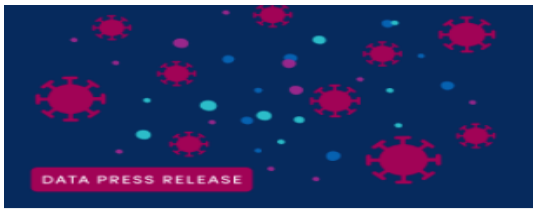

Fig. 2(b). One of the many items of additional information provided on the website integrated with the app

Reassurance about anonymized data is provided by comprehensive explanation to build trust and this may be reinforced by a university brand, thus the negative privacy fear may be counter-balanced by positive perceptions of honesty, responsibility, and integrity. The King's/Zoe app also provides a wide variety of information about Covid-19 research, local infection risk data, advice on personal behavior to avoid infection, etc.

Over 4 million users have downloaded the app, representing ~10-15% of the overall potential user population. Although this is a significant success for a product with no advertising and awareness spread primarily by word of mouth, the cost/effort of use for most people may not outweigh the altruism/helpfulness benefits in spite of the wealth of useful information on the website [41]. The King's/Zoe app may be hindered by the cost of continuing use. Daily logging of one's personal health condition is an operational chore and a memory burden which is offset against the personal helpfulness benefit and the altruistic motivation to contribute to scientific research. As the information-related benefits can be acquired without using the symptom tracker, users can 'free load' on the efforts of others, which may tip the balance against prolonged use.

## CONCLUSIONS AND LESSONS LEARNED

Although the conclusions of our value-feature analysis were consistent with the history of the NHS app versions, it is apparent that values alone do not provide sufficient insight into questions of user acceptance. All experience reports highlight security and data privacy concerns [42]; however, download success varies considerably between nations and all are below the necessary 60% acceptance for effective Covid-19 infection control [43]. Success in reducing the Covid-19 infection rate (R) has only been apparent in South Korea and China, which both have mandatory use and centralized models. In these cases, authority replaces choice and the influence of users' values [44]. Germany has experienced more success with a decentralized model with 18 million downloads, ~ 25-30% of the potential use base; however, this is still below the 60% threshold for effective infection control. Furthermore, critics have noted that the benefits of the app (i.e. helpfulness) have not been explained.[1] Several European nations have adopted the decentralized model, consistent with our values analysis; however, no successful implementation approaching 50% downloading has been reported to date.

To understand download success, investigation into user costs and motivations is essential; for example, the costs (effort) of downloading, installing, learning to use and then actually operating the application. Motivations are more likely to be more closely related to values, for instance security in terms of mitigating the perceived danger of catching Covid, as well as security in terms of privacy concerns over insecure data. Clearly, contextual interpretation of values will be important for understanding the relationship to functionality. Furthermore, consideration of potential emotional reactions (i.e. fear and anxiety for security/privacy data loss threats) is a necessary component of a cost-benefit analysis. Personal costs of use, both economic purchase/subscriptions and, more importantly, effort of use, needs to be balanced against personal (i.e. achievement, helpfulness) and societal (e.g. altruism) benefits. The cost-benefit analysis of the apps is summarized in Table 3.

Table 3. Cost-benefit analysis for the first NHS Apps versions (vn1,vn2) and King's/Zoe Covid-19 apps.
– denotes negative score of costs (*italic*), while + represents a benefit score, with the net figure as a simple sum of positive and negative scores.

|  | NHS V1 | NHS V2 | King's/Zoe App |
|---|---|---|---|
| *Download* | -- | -- | -- |
| *Use* | -- | -- | -- -- |
| *Data loss* | -- -- | -- | -- -- |
| Personal | + | + | + + |
| Society | + | + | + + |
| Net | -2 | -1 | -1 |

For the NHS apps (vn1 and vn2), once the user has spent time downloading and registering themselves with the app, there are no operational costs until the app detects proximity to an infected person or the user becomes infected themselves. In the latter case there is a self-reporting cost. In both cases there is the social cost of self-isolation. The King's/Zoe app does not have automatic proximity detecting so it imposes a daily operational self-reporting cost. Set against these costs are similar personal benefits of being alerted to the potential danger of infection, with the additional information/advice and societal benefits of contributing to Covid-19 research in the King's/Zoe app [41]. The net benefit estimate indicates there is little to choose between the apps, although centralized architectures tend to impose more privacy fears (data loss) which could be countered by positive perceptions of society and personal benefit (i.e. helpfulness value). Individual choice may depend on perception of the effectiveness of the apps for avoiding Covid-19 infection.

---

[1] https://www.dw.com/en/too-few-germans-using-coronavirus-pandemic-tracing-app/a-54970227

Benefits in many systems will tend to be intangible rather than economic and may be assessed in surveys in terms of contributions to users' values. Costs, in contrast, tend to be physical in terms of learning and operating systems and these costs may be incurred in advance of benefits; hence the motivations underpinning benefits need to be communicated clearly to users to ensure they overcome the cost barriers before benefits arise.

Value-feature analysis can provide insight into potential users' response to software design; however, evaluating users' value-oriented judgements could consume considerable effort in completing questions and focus group interviews. However, users' surveys will be necessary for accurate data, for example the value assignments we made in the case study will actually depend on users' awareness of features, desire for absent features, and actual experience with the product. Proactive design advice linked to value analysis in the requirements phase may have more immediate utility.

## DISCUSSION

Our contribution in this preliminary investigation has been to articulate the potential impact of values on design features (RQ1) and provide some evidence that values alone are insufficient for understanding product success (RQ2). The case study provides evidence for the forensic potential of values used as a critical lens through which features of an application, and possible reasons for success or failure of a product, may be questioned. This approach is similar to [2]'s critical deconstruction of GDPR legislation to illustrate possible impacts of legal rights and principles on values. However, value-feature analysis on its own did not show a strong differentiation between the Covid-19 apps; furthermore, it did not indicate the poor success rate of Covid-19 tracing apps in several countries. We believe the answer to the question posed in RQ2 is that user motivation and costs need to be assessed as well as values.

Previous research road maps for values in SE [3-4] have argued for the development of value-based analysis methods and measurement of values [45]. Although methods for value-based SE have been proposed [2], few reports of industrial application have emerged. One exception is Value-Based Requirements Engineering (VBRE) [8] which incorporated taxonomies of values, motivations and emotions in a method that was validated by case study use in industry. The only other value-based method with industrial validation is VSD which has an extensive record of application [7, 46]. Further case studies of applications of value in SE within industrial contexts are needed.

We argue that value analysis in SE has two potential contributions. First is when the overall system purpose has a strong value orientation; for instance, in web applications to promote sustainability, awareness of climate change and green values. Other examples may be found in healthcare applications intended to ameliorate mental illness and loneliness with empathy, helpfulness, and support; and more generally in persuasive technology [47] and recommender systems that pose trade-off challenges between benefits in terms of satisfying user goals with achievement values, set against intrusiveness with negative connotations for autonomy, and fairness. Fairness and justice values will be important, especially when decisions are powered by big data and machine learning, where the outcomes need to be seen to be fair, transparent, and just [33, 48]. Values should motivate exploring the 'horizon of possibilities' in such systems as principles and heuristics informing design. For example, when creativity and helpfulness values are identified, system goal and non-functional requirements of flexibility/ adaptability could focus analysis on requirements for configuration, personalization, and end-user development [25, 49].

The second contribution lies in requirements validation and modeling where values provide the principles and standards by which system action and performance are assessed. This needs to develop from meta-models that facilitate mapping values to soft goals in i* strategic dependency models [20] to more active guidance about value-based requirements implications. The 'envelope of acceptability' could specify the range of actions that a system can enact in the world, with values and related NFRs dictating which actions should be permissible, extending the concept of system monitors and awareness requirements [30, 32, 50-51]. Robotics is one example of ethical and safety/security values in action, where system action must not endanger human operators [10, 33].

However, values are insufficient for analyzing the reasons for systems failure. For example, security and privacy concerns are obvious value implications in Covid-19 track-and-trace apps; however, these values alone do not provide insight into the variation in success between different app designs [37, 44]. While privacy/security is a frequently reported value-related concern, other issues such as compatibility with mobile phone OS, battery life, and accuracy of contact predictions have been barriers cited in surveys of Covid app success [35, 52]. Trade-off analysis [e.g. 53] between motivations, convenience and costs should be considered, as illustrated in a survey of app downloading in Australia [54]. Our future research will expand the ambit of values to include constructs and lessons learned from technology acceptance modeling [11, 12], as well as developing further guidance on design implications to improve user benefits by value-aware feature analysis.

## ACKNOWLEDGEMENTS

This research was partially supported by EPSRC project (EP/T017627/1) Twenty-20 Insight.

REFERENCES

[1] B. Penzenstadler, L. Duboc, C. C. Venters, et al., "Software engineering for sustainability: Find the leverage points!", IEEE Software, 2018, pp. 22-33.

[2] H. Perera, W. Hussain, D. Mougouei, R. A. Shams, A. Nurwidyantoro, and J. Whittle, "Towards integrating human values into software: Mapping principles and rights of GDPR to values," in 2019 IEEE 27th International Requirements Engineering Conference, IEEE, 2019, pp. 404-409.

[3] D. Mougouei, H. Perera, W. Hussain, R. Shams, and J. Whittle, "Operationalizing human values in software: A research roadmap," in ESEC/FSE'18 - Proceedings of the 26th ACM Joint Meeting on European Software Engineering Conference, 2018, pp.780-784.

[4] J. Whittle, M. A. Ferrario, W. Simm, W. and W. Hussain, "A case for human values in software engineering", IEEE Software, 2019, 11 p.

[5] I. Sommerville, T. Rodden, P. Sawyer, R. Bentley, and M. Twidale, "Integrating ethnography into the requirements engineering process", in Proceedings IEEE Symposium on Requirements Engineering, 1992.

[6] M. Jirotka, R. Proctor, M. Hartswood, et al., "Collaboration and trust in healthcare innovation: The eDiaMoND case study", Computer Supported Cooperative Work (CSCW), 2005, pp. 369-398.

[7] B. Friedman, "Value sensitive design", in D. Schular (ed.) Liberating Voices: A Pattern Language for Communication Revolution. MIT Press, Cambridge MA, 2008.

[8] S. Thew and A. G. Sutcliffe, "Value based requirements engineering: method and experience," Requirements Engineering, 2018, pp. 443-464. https://doi.org/10.1007/s00766-017-0273-y

[9] S. Thew, Value-based Requirements Engineering. PhD Thesis, University of Manchester, 2015.

[10] A. Paiva, I. Leite, H. Boukricha, and I. Wachsmuth, "Empathy in virtual agents and robots: A survey," TiiS, 2017, pp. 11:1-11:40.

[11] N. Marangunić, and A. Granić, "Technology Acceptance Model: A literature review from 1986 to 2013", Universal Access in the Information Society, 2015, pp. 81-95.

[12] V. Venkatesh, J. Y. L. Thong, and X. Xu, "Consumer Acceptance and Use of Information Technology: Extending the Unified Theory of Acceptance and Use of Technology", MIS Quarterly, 2012, pp. 157-178.

[13] I. Ramos and D. M. Berry, "Is emotion relevant to requirements engineering?", Requirements Engineering, 2005, pp. 238-242.

[14] R. Fuentes-Fernandez, J. Gomez-Sanz, and J. Pavon, "Understanding the human context in requirements elicitation", Requirements Engineering 2010, pp. 267-283.

[15] J. Viega, T. Kohno, and B.Potter, "Trust (and mistrust) in secure applications", Communications ACM, 2001, pp. 31-36.

[16] M. Krumbholz, N. A. M. Maiden, B. Wangler, and L. Bergman, "How culture might impact on the implementation of Enterprise Resource Planning packages", in Advanced Information Systems Engineering (1789). Springer, Berlin, 2000, pp 279-293.

[17] G. Cockton, "Worth-focused design", In J. M. Carroll (Ed.), Balance, Integration, and Generosity. (Morgan Claypool HCI series), 2020.

[18] S. H. Schwartz, "A theory of cultural values and some implications for work," Applied Psychology, 1999, pp. 23-47.

[19] M. B. Holbrook, Consumer Value: A Framework for Analysis and Research. Routledge, London, 1998.

[20] J. Zdravkovic, E. O. Svee, and C. Giannoulis, C. "Capturing consumer preferences as requirements for software product lines", Requirements Engineering, 2015, pp.71-90.

[21] A. S. Cheng, and K. R. Fleischmann, "Developing a meta-inventory of human values," in Proceedings of the ASIST Annual Meeting, 2010.

[22] A. G. Sutcliffe, "An architecture framework for self-aware adaptive systems", in I. Mistrík et al., Economics-Driven Software Architecture (EDSA), Elsevier, 2014.

[23] E. Mumford, Values, Technology and Work. Martinus Nijhoff, Amsterdam, 1981.

[24] E. Hollnagel, and A. Bye, "Principles for modelling function allocation", International Journal of Human-Computer Studies, 2000, pp. 253-265.

[25] G. Fischer, Domain-Oriented Design Environments. Automated Software Engineering, 1994, pp. 177-203.

[26] E. Yu, Social Modeling for Requirements Engineering Cooperative Information Systems. MIT Press, Cambridge MA, 2010.

[27] A. G. Sutcliffe, Analysing the effectiveness of human activity systems with i*. In E.Yu et al. (eds), Social Modelling for Requirements Engineering. MIT Press, Cambridge MA, 2011.

[28] B. Hui, S. Laiskos, and J. Mylopoulos, "Requirements analysis for customisable software: A goals skills preferences framework. In Proceedings IEEE Joint International Conference on Requirements Engineering. IEEE Computer Society Press, Los Alamitos CA, 2003.

[29] A. G. Sutcliffe, "Designing for user engagement: Aesthetic and attractive user interfaces", In J. M. Carroll (ed.) Synthesis Lectures on Human Centred Informatics. Morgan and Claypool, 2009.

[30] V. E. S. Souza, A. Lapouchnian, W. N. Robinson and J. Mylopoulos, "Awareness requirements", in Software Engineering for Self-Adaptive Systems II (pp. 133-161). Springer, Berlin, 2013.

[31] A. G. Sutcliffe and P. Sawyer, "Requirements elicitation: Towards the unknown unknowns", in Proceedings of 21st IEEE International Requirements Engineering Conference, 2013, pp. 92-104.

[32] F. Dalpiaz, P. Giorgini, and J. Mylopoulos, "Adaptive socio-technical systems: A requirements-based approach", Requirements Engineering, 2013, pp. 1-24. doi: 10.1007/s00766-011-0132-1.

[33] R. K. E. Bellamy, K. Dey, M. Hind, S. C. Hoffman, et al. "Think your artificial intelligence software is fair? Think again", IEEE Software, 2019, pp. 76-80.

[34] L. Ferretti, C. Wymant, M. Kendall, et al., "Quantifying SARS-CoV-2 transmission suggests epidemic control with digital contact tracing", Science, 2020, 368, 619.

[35] N. Ahmed, R. A. Michelin, W. Xue, S. Ruj et al., "A survey of Covid-19 contact tracing apps", IEEE Access, 2020, pp.134577-134601.

[36] BCS. The NHS contact tracing app: Nine key talking points. 2020. https://www.bcs.org/content-hub/the-nhs-contact-tracing-app-9-key-talking-points/

[37] R. Cellan-Jones, and L. Kelion, "Coronavirus: The great contact-tracing apps mystery", BBC News, https://www.bbc.com/news/technology-53485569, 2020.
[38] NHS, "COVID-19 app support", https://covid19.nhs.uk, 2020.
[39] Wired, "What's really happening with the NHS Covid-19 app trial?" https://www.wired.co.uk/article/contact-tracing-app-isle-of-wight-trial, 2020.
[40] R. Cellan-Jones and Z. Kleinman, "NHS Covid-19 app: 12m downloads - and lots of questions BBC News, https://www.bbc.co.uk/news/technology-54326267, 2020.
[41] Zoe, "About, news and research", https://covid.joinzoe.com/data, 2020.
[42] O. Pereira, "Why should we install the Coronalert contact tracing app?" Technical Report Crypto group, UCLouvain, Belgium, 2020.
[43] M. E. Kretzschmar, G. Rozhnova, M. C. Bootsma et al., "Impact of delays on effectiveness of contact tracing strategies for COVID-19: A modelling study", The Lancet Public Health, 2020, pp. e452-e459.
[44] A. Urbaczewski, and Y. J. Lee, Y. J., "Information technology and the pandemic: A preliminary multinational analysis of the impact of mobile tracking technology on the COVID-19 contagion control", European Journal of Information Systems, 2020. doi: 10.1080/0960085X.2020.1802358
[45] E. Winter, S. Forshaw, and M. A. Ferrario, "Measuring human values in software engineering," in Proceedings of the 12th ACM/IEEE International Symposium on Empirical Software Engineering and Measurement. ACM, 2018, p. 48.
[46] B. Friedman, D. Hendry, and A. Borning, A Survey of Value Sensitive Design Methods. (Foundations and Trends in Human Computer Interaction), Now Publishers, Boston and Delft, 2017.
[47] B. J. Fogg, Persuasive Technology: Using Computers to Change What We Think and Do. San Francisco: Morgan Kaufmann, 2003.
[48] R. Guidotti, A. Monreale, S. Ruggieri, F. Turini, F. Giannotti, and D. Pedreschi, "A survey of methods for explaining black box models", ACM Computing Surveys, 2018, p. 93.
[49] G. Fischer, E. Giaccardi, Y. Ye, A. G. Sutcliffe, and N. Mehandjiev, "A framework for end-user development: Socio-technical perspectives and meta-design", Communications of the ACM, 2004, pp. 33-39.
[50] P. Sawyer, N. Bencomo, J. Whittle, E. Letier, A. Finkelstein, "Requirements-Aware Systems A research agenda for RE for self-adaptive systems", RE2010, Australia, 2010.
[51] N. Bencomo, J. Whittle, P. Sawyer, A. Finkelstein, and E. Letier, "Requirements Reflection: Requirements as Runtime Entities", ICSE 2010, Track NIER, South Africa, 2010.
[52] R. Abbas, and K. Michael, K., 2020. "COVID-19 contact trace app deployments: Learnings from Australia and Singapore", IEEE Consumer Electronics Magazine, 2020, pp. 65-70.
[53] B. Boehm, and H. In, "Identifying quality-requirement conflicts", IEEE Software, 1996, pp. 25-35.
[54] Conversation, "70% of people surveyed ...", https://theconversation.com/70-of-people-surveyed-said-theyd-download-a-coronavirus-app-only-44-did-why-the-gap-138427, 2020.